\documentclass[12pt]{article}
\usepackage{graphicx}
\usepackage{hyperref}

\newcommand\pubnumber{SNSN-323-63}
\newcommand\pubdate{\today}

\def\institute{Universidad de Oviedo, SPAIN}

\def\Title#1{\begin{center} {\Large #1 } \end{center}}
\def\Author#1{\begin{center}{ \sc #1} \end{center}}
\def\Address#1{\begin{center}{ \it #1} \end{center}}

\newcommand\pubblock{\rightline{\begin{tabular}{l} \pubnumber\\
         \pubdate  \end{tabular}}}
\newenvironment{Abstract}{\begin{quotation}  }{\end{quotation}}
\newenvironment{Presented}{\begin{quotation} \begin{center} 
             PRESENTED AT\end{center}\bigskip 
      \begin{center}\begin{large}}{\end{large}\end{center} \end{quotation}}

\newcommand{\finalxsec}{
  \begin{displaymath}
  \sigma_{t\bar{t}}^{pp}(5.02~TeV) = 69.5 \pm 6.1 (stat) \pm 5.6 (syst) \pm 1.6 (lumi)~pb = 69.5 \pm 8.4 (total)~pb
  \end{displaymath}
}

\newcommand{\xsecth}{
  \begin{displaymath}
  \sigma_{t\bar{t}}^{NNLO} = 68.9 \pm 1.9 (scale) \pm 2.3 (PDF) \pm 1.4 (\alpha_{S})~pb
  \end{displaymath}
}

\newcommand{\ttbar}{$\textrm{t}\bar{\textrm{t}}$~}

\newcommand{\sqrts}{$\sqrt{s}$ = ~}
\newcommand{\met}{$p_{T}^{miss}~$}
\newcommand{\ljets}{$\ell$+jets }
\newcommand{\pt}{$p_{T}~$}
\newcommand{\abseta}{$\vert \eta \vert~$}
\newcommand{\stt}{$\sigma_{t\bar{t}}~$}
\newcommand{\arxiv}[1]{\href{https://arxiv.org/abs/#1}{arXiv:#1}}

\def\beq{\begin{equation}}
\def\eeq#1{\label{#1}\end{equation}}
\def\eeqn{\end{equation}}

\def\beqa{\begin{eqnarray}}
\def\eeqa#1{\label{#1}\end{eqnarray}}
\def\eeqan{\end{eqnarray}}

\let\bar=\overbar

\def\Dslash{\not{\hbox{\kern-4pt $D$}}}
\def\dslash{\not{\hbox{\kern-2pt $\del$}}}

\def\msb{{\bar{\ssstyle M \kern -1pt S}}}

\begin{document}
\begin{titlepage}
\pubblock

\vfill
\Title{Measurement of the \ttbar inclusive cross section at \sqrts 5.02 TeV in pp collisions with CMS}
\vfill
\Author{Juan R. Gonz{\'a}lez Fern{\'a}ndez on behalf of the CMS Collaboration}
\Address{\institute}
\vfill
\begin{Abstract}
The top quark pair inclusive production cross section (\stt) is measured in pp collisions at a centre-of-mass energy of 5.02 TeV. The analyzed data have been collected by the CMS experiment at the CERN LHC and correspond to an integrated luminosity of 27.4 /pb. The measurement is performed by analyzing events with at least one charged lepton. The measured cross section is 69.5 $\pm$ 8.4 pb. The result is in agreement with the expectation from the standard model. The impact of the presented measurement on the gluon distribution function is illustrated through a quantum chromodynamic analysis at next-to-next-to-leading order.
\end{Abstract}
\vfill
\begin{Presented}
$10^{th}$ International Workshop on Top Quark Physics\\
Braga, Portugal,  September 17--22, 2017
\end{Presented}
\vfill
\end{titlepage}
\def\thefootnote{\fnsymbol{footnote}}
\setcounter{footnote}{0}

\section{Introduction}
The study of the top quark, the heaviest elementary particle of the SM, is crucial to understand several questions in QCD. In proton-proton collisions, these quarks are mostly produced in pairs by gluons fusion. The inclusive producction cross section of top quark pairs (\stt) has been measured at different centre-of-mass energies in pp collisions by the CMS \cite{bib:cms} and ATLAS collaborations, probing QCD predictions and constraining new physics scenarios, proton parton distribution functions, the top quark pole mass, $\alpha_{S}$, and other quantities.

In November 2015, the LHC delivered pp collisions at \sqrts 5.02 TeV. At this centre-of-mass energy, top quark pairs are produced by high-x gluons, so a measurement of \stt can be used to constrain gluon PDFs. In this document a summary of the recent measurement of \stt at 5.02 TeV using events with at least one charged lepton is presented \cite{bib:TOP16023}. The SM prediction for the \ttbar cross section is used to normalize the expected signal yields and has a value of \cite{bib:thxsec}:
\xsecth

The analysis is divided in two different final states: \ljets channel, where one W boson decays hadronically, giving two jets in the final state, and the other decays into one light charged lepton (e, $\mu$) and a neutrino, and dilepton channel, where both W bosons decay into leptons. In the former case, \stt is extracted by a fit to the distribution of a kinematic variable for different categories of lepton flavor and jet multiplicity, while in the dilepton case an event counting technique is used.

\section{Object reconstruction and event selection}
The particle-flow (PF) algorithm \cite{bib:pf} is used to reconstruct individual particles, using the combined information from all CMS sub-detectors. The reconstruction and identification efficiencies for the leptons are measured using a tag-and-probe technique and the simulation is corrected using \pt and \abseta dependent scale factors.

Jets are reconstructed from PF candidates using a cone of $\Delta R = $ 0.4. A jet can be identified as coming from a b quark (b-tagged) using a combined secondary vertex algorithm \cite{bib:btagging}.

The missing transverse momentum vector (\met) is reconstructed as the negative vector sum of the momentum of all PF candidates in an event, projected into the transverse plane.

For the \ljets channel, muons (electrons) with \pt $\geq$ 25 (40) GeV and \abseta $\leq$ 2.1 (2.5) are considered. Events with at least two light jets with \pt $\geq$ 30 GeV and \abseta $\leq$ 2.4 are selected and classified into b-tag multiplicity categories (0b, 1b, $\geq$2b). Events are rejected in this channel if there is an extra electron or muon with \pt $\geq$ 10 GeV and looser identification criteria.

Dilepton events are required to have at least one muon candidate with \pt $\geq$ 18 GeV and \abseta $\leq$ 2.1 and at least two jets with \pt $\geq$ 25 GeV and \abseta $\leq$ 3.0. Electrons are required to have \pt $\geq$ 20 GeV and \abseta $\leq$ 2.4. The selected events must have a dilepton invariant mass of $M_{\ell\ell} \geq$ 20 GeV. For events containing two muons, in order to suppress Drell-Yan (DY) background, the invariant mass of the dimuon pair is required not to be in the range 76 yo 106 GeV and the event must have a missing transverse energy of at least 35 GeV.

\section{Background estimate}
In the \ljets channel the QCD multijet background is estimated from data in a dedicated region where the muon (electron) fails the isolation (identification) criteria. All other backgrounds are taken from Monte Carlo simulation. The normalization of QCD multijet background is more accurately obtained in the fit.

In the dilepton final state DY background normalization is obtained in a control region using events with same flavor and an invariant mass in the range of 76 to 106 GeV. Background contamination from events containing a fake lepton (mostly W+Jets and semileptonic $t\bar{t}$) is estimated from a control region with similar event selection or looser isolation criteria but same-sign leptons. All other backgrounds are estimated from simulation.

\section{Results and QCD analysis}
\begin{figure}[htb]
\centering
\includegraphics[height=1.5in]{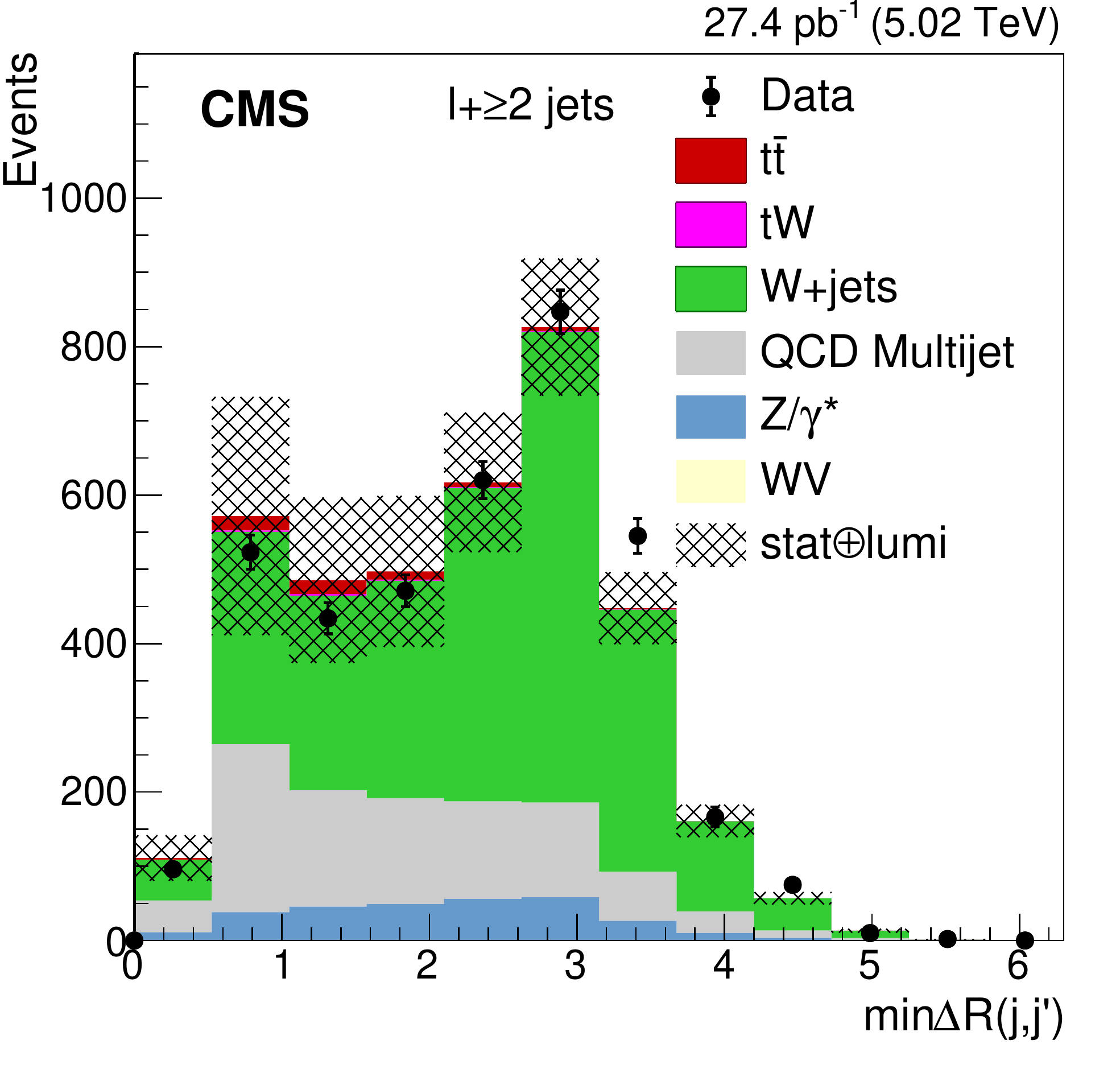}
\includegraphics[height=1.5in]{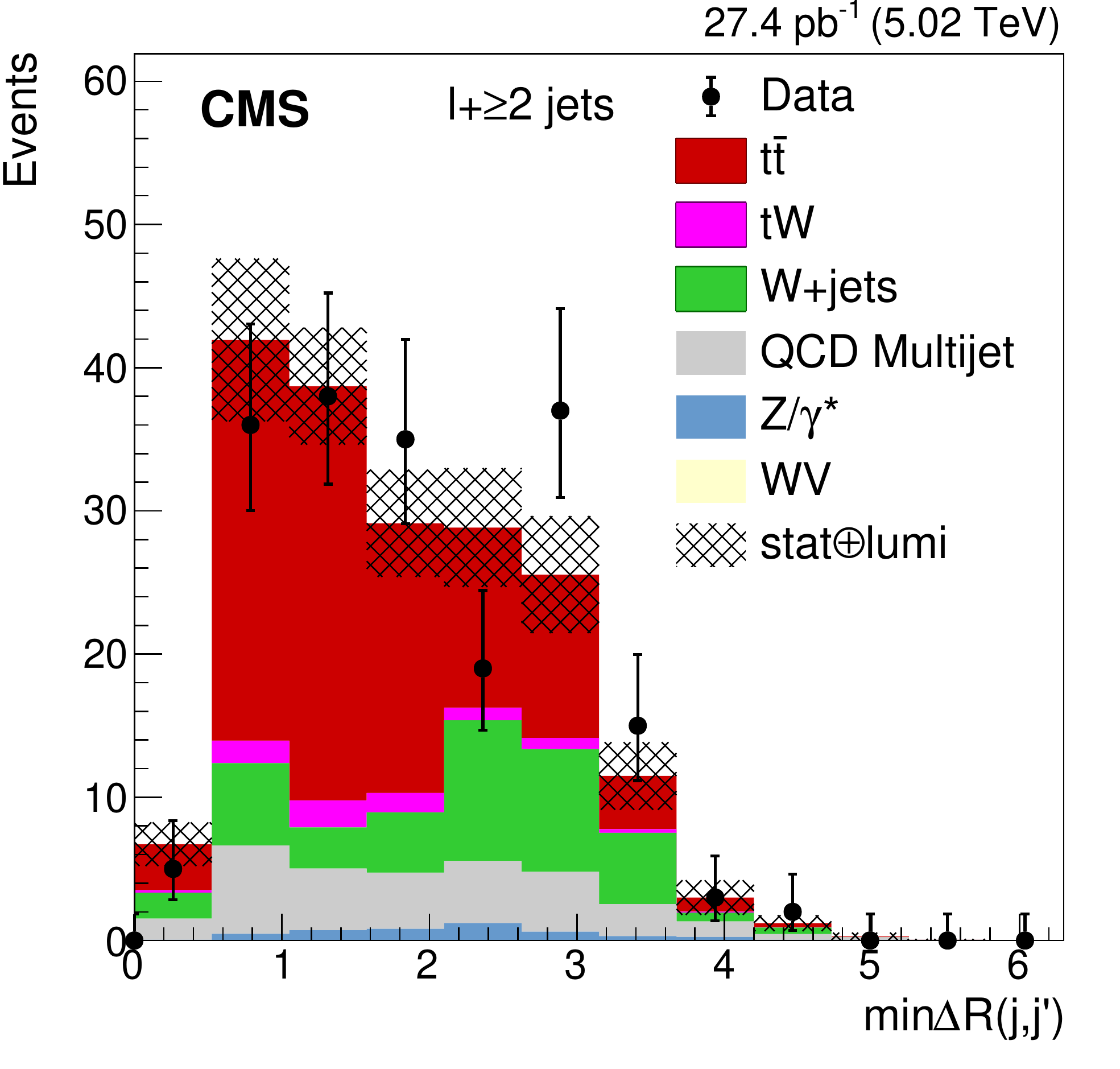}
\includegraphics[height=1.5in]{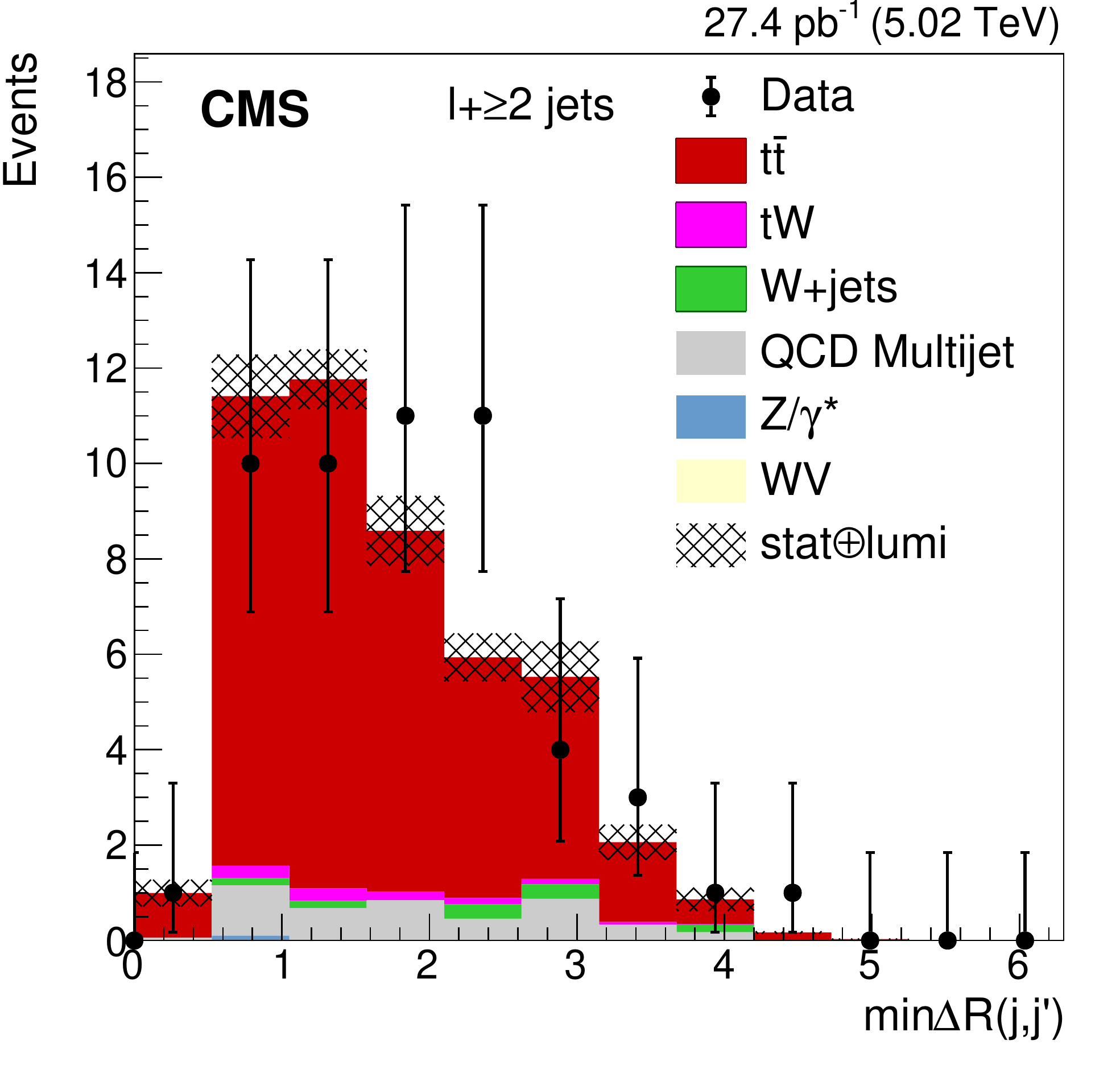}
\caption{Minimum $\Delta R(j,j')$ in the event in each category of b-tag multiplicity. These distributions are fitted to extract \stt \cite{bib:TOP16023}.}
\label{fig:alldrjj}
\end{figure}

In the Figure \ref{fig:alldrjj} the minimum $\Delta R$ distance between all pair of jets in the event is shown for data and prediction in the \ljets channel for all the b-tag multiplicity categories. These distributions are fitted to extract \stt. Also, the QCD background normalization is adjusted in the fit, the b-tagging efficiency is extracted and several systematic uncertainties are used as nuisances in the fit.

\begin{figure}[htb]
\centering
\includegraphics[height=1.5in]{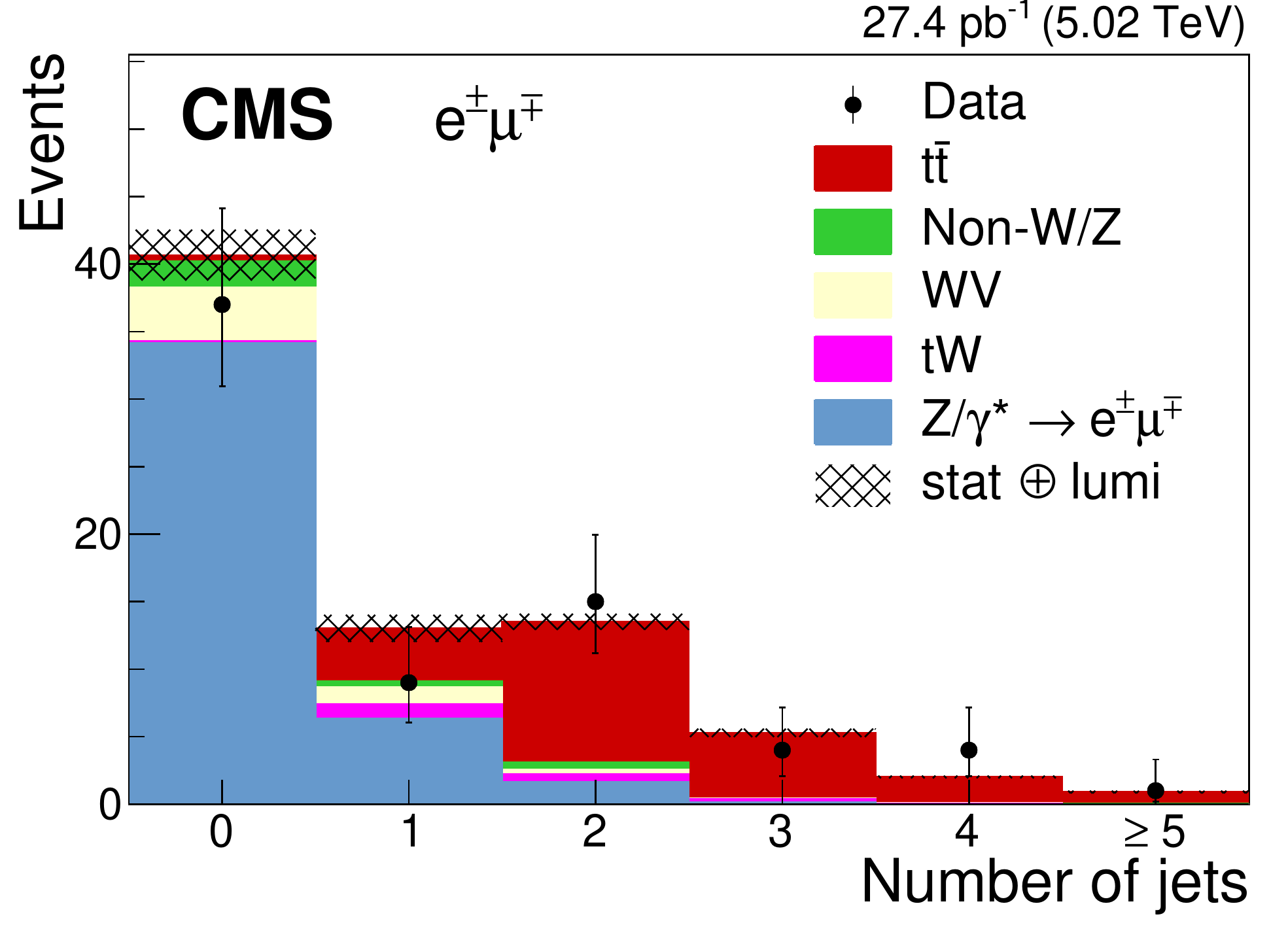} \hspace{1cm}
\includegraphics[height=1.5in]{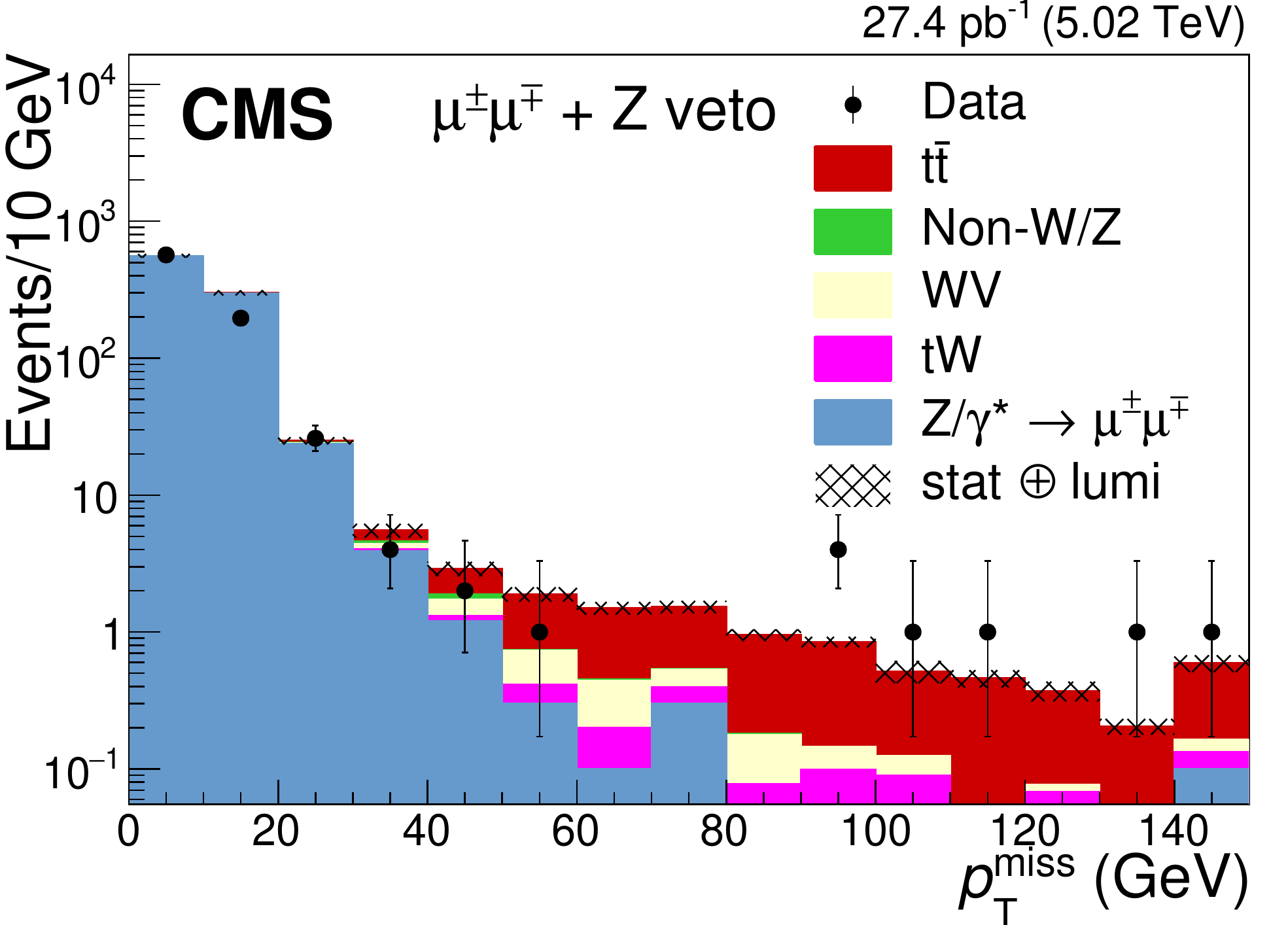}
\caption{Jet multiplicity for events with an $e\mu$ pair (left) and \met distribution for events with a pair of opposite-sign muons (right) \cite{bib:TOP16023}.}
\label{fig:dilepfigs}
\end{figure}

 In the Figure \ref{fig:dilepfigs} the jet multiplicity for eventas containing an $e\mu$ pair and the \met for events containing a dimuon pair are shown. After cutting on these variables, the cross section is calculated by counting the number of observed data, subtracting the background prediction and extrapolated to the full phase space. 

The statistical uncertainty is dominant in all the final states. The measurements in the different channels are combined, obtaining a final measurement of the \ttbar inclusive cross section at 5.02 TeV of:
\finalxsec
in agreement with the SM prediction and with a total uncertainty of 12\%.

\begin{figure}[htb]
\centering
\includegraphics[height=1.5in]{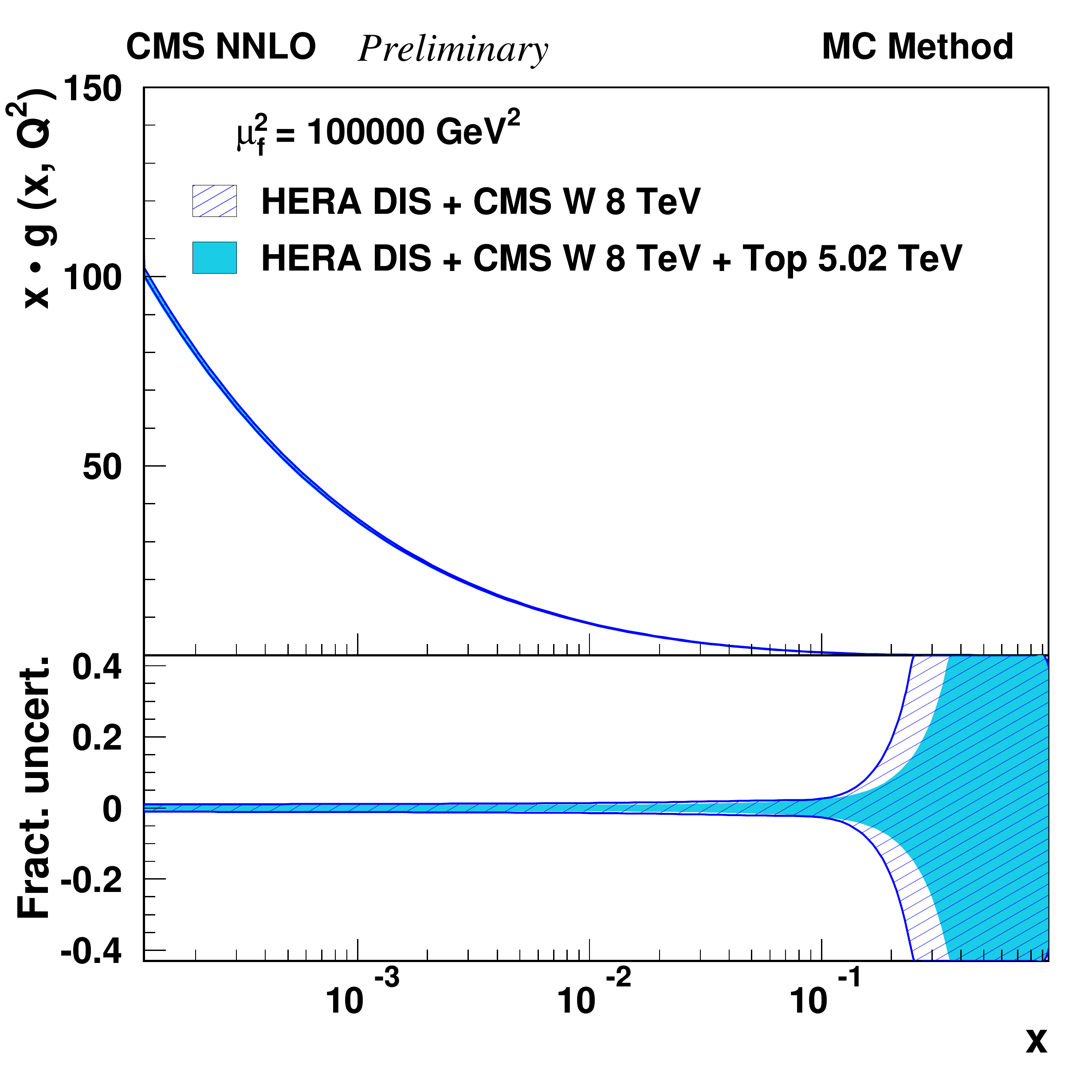} 
\caption{The gluon distribution function and its uncertainty are shown as a function of x. The effect of the \stt measurement on the result can be seen comparing the uncertainty from a QCD analysis using the HERA DIS and CMS muon charge asymmetry measurements (hatched area), and also including the \stt result (solid area) \cite{bib:TOP16023}.}
\label{fig:gluonpdf}
\end{figure}

This result has been used to perform a QCD analysis combining this measurement with data from HERA DIS and CMS measurement of muon charge asymmetry. In Figure \ref{fig:gluonpdf} the result of the QCD analysis for the gluon distribution function in the proton is shown.

\section{Conclusions}
\begin{figure}[htb]
\centering
\includegraphics[height=1.5in]{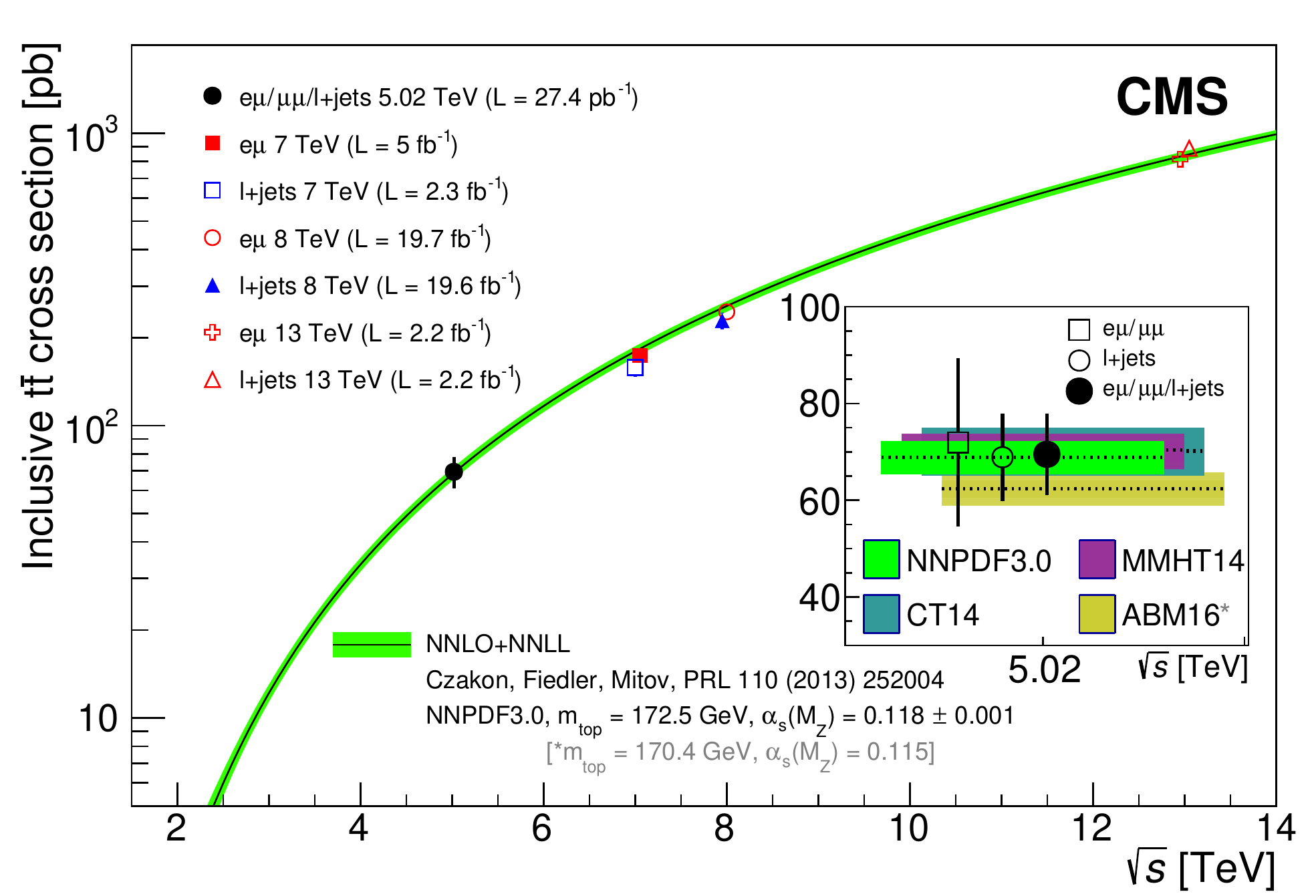} 
\caption{Summary of \ttbar cross section measurements at different centre-of-mass energies at CMS \cite{bib:TOP16023}.}
\label{fig:xsecsqrts}
\end{figure}

The first measurement of the \ttbar inclusive production cross section at 5.02 TeV is summarized in this document. The result is in agreement with the SM prediction and has a total uncertainty of 12\%. This result means a new point in the $\sigma_{t\bar{t}}(\sqrt{s})$ curve, shown in Figure \ref{fig:xsecsqrts}, which summarizes a great piece of our current knowledge of QCD. 

The measurement has been proven to be useful to probe high-x gluon in proton PDF through a QCD analysis, so this can lead to new measurements of \stt at this center-of-mass energy, with more data and smaller uncertainties, that can give a crucial contribution to further constrain proton PDF.

\end{document}